\documentclass[a4paper,11pt]{article}

\usepackage{soul}
\usepackage{color}
\usepackage[singlelinecheck=off,center]{caption}
\usepackage[T2A]{fontenc}
\usepackage[utf8x]{inputenc}
\usepackage[warn]{mathtext}
\usepackage{indentfirst}
\usepackage[fleqn]{amsmath}
\usepackage{amsthm,amsfonts,amssymb,amscd}
\usepackage{hyperref}
\usepackage{bm}
\usepackage{icomma}
\usepackage{mathrsfs}
\usepackage{cite}

\usepackage{physics}

\begin{document}

\title{The generalized uncertainty principle within the ordinary framework of quantum mechanics}
\author{Y. V. Przhiyalkovskiy \\ \\ \textit{\small Kotelnikov Institute of Radioengineering and Electronics (Fryazino Branch)}\\ \textit{\small of Russian Academy of Sciences,}\\ \textit{\small Vvedenskogo sq., 1, Fryazino, Moscow reg., 141190, Russia} \\ \small e-mail: yankus.p@gmail.com}
\date{}

\title{The generalized uncertainty principle within the ordinary framework of quantum mechanics}%

%
%

\maketitle

\begin{abstract}

A proper deformation of the underlying coordinate and momentum commutation relations in quantum mechanics provides a phenomenological approach to account for the influence of gravity on small scales. Introducing the squared momentum term results in a generalized uncertainty principle, which limits the minimum uncertainty in particle position to the Planck length. However, such a deformation of the commutator significantly changes the formalism, making it separate from the canonical formalism of quantum mechanics. In this study, it is shown that the deformed algebra of position and momentum operators can be incorporated into the framework of ordinary quantum mechanics.

\end{abstract}


\section{Introduction}

It is known, that quantum theory faces a fundamental challenge at Planck scales when gravity is taken into account. The arguments that clearly illustrate this follow from a thought experiment in which the position of a particle is measured by photon scattering~\cite{Plato2016}. The resolution we can reach in such a measurement is limited by the wavelength, which is inversely proportional to the energy. Therefore, to improve the precision, additional energy must be pumped into the photons. In general, quantum mechanics itself does not prohibit measuring the position with any desired accuracy. But as we approach the Planck scale $\sim \!\! 10^{-33}$~cm in measuring resolution, the energy of photons becomes so high that, according to general relativity, they significantly perturb the surrounding space. This causes the particle to move additionally during the interaction. As a result, the further shortening of the photon wavelength beyond this threshold will degrade the accuracy of the position measurement.

Although a unified theory of quantum gravity has not yet been formulated, there are several approaches that can be applied in quantum mechanics to address the problem of small scales. String theory, for example, assigns a small but finite size to any particle~\cite{Amati1989}. As a result, there is a natural limit below which the very notion of position becomes irrelevant. Essentially the same applies to quantum loop gravity \cite{Rovelli1995} and theories with non-commutative geometry \cite{Lawson2020}. Another way to introduce a finite resolution into the quantum mechanical framework, which has received considerable attention in the literature, is to deform the underlying canonical commutation relations \cite{Maggiore1993, Kempf1995}. It turns out that the appropriate deformation of the position and momentum commutator, while not changing much on a large scale, excludes the existence of quantum mechanical states with position uncertainty less than a certain value. It is this quantity that is interpreted as the minimum length. Formally, the minimal uncertainty is a manifestation of what is known as the generalized uncertainty principle (GUP).

Quantum theory based on deformed commutation relations has been extensively studied since the early 1990s~\cite{Kempf1995}. It was not only a promising approach to such fundamental problems as the unification of quantum mechanics and general relativity, but also found applications in ultraviolet and infrared regularization \cite{Kempf1997, Kempf1996}, the study of black holes~\cite{Gangopadhyay2014, Faizal2015}, thermodynamics \cite{Bernardo2018}. It was shown that deforming the commutation relations introduces specific peculiarities into the mathematical framework that must be taken into account \cite{Kempf1994}. The deformed algebra of position and momentum operators inevitably differ from its canonical counterparts used in ordinary quantum mechanics. In particular, the position operator becomes exclusively symmetric in the domain of physical states. The crucial point here is that the eigenvectors of the position operator are not physical states. Strictly speaking, this is also the case in ordinary quantum mechanics, but unlike the latter, we can't even approach the eigenvectors through a set of physical states.
Moreover, in order for the position operator to be symmetric, one is forced to modify the Hilbert space by introducing a measure into the inner product. The constructed formalism proves to be self-consistent, i.e. isolated from the canonical one, and reduces to the latter only in the limit. As a result, it becomes difficult to compare the two formalisms. The main idea behind the present work is to implement the GUP into the ordinary formalism of quantum mechanics. For this, we construct a representation of the deformed operator algebra in the canonical Hilbert space. Having done that, we can consider quantum theory with the GUP as just a part of ordinary quantum mechanics.

\section{Generalization of the uncertainty principle}

In the most popular generalization of the Heisenberg uncertainty principle, a non-zero minimum length in non-relativistic quantum theory could be obtained by inserting the quadratic term for the momentum uncertainty~\cite{Kempf1995}:
\begin{equation}
	\Delta X \Delta P \ge \frac{\hbar}{2} \big( 1 + \beta \Delta P^2 \big),
\end{equation}
where $\Delta X = \sqrt{\langle \hat{X}^2 \rangle - \langle \hat{X} \rangle^2}$ and $\Delta P = \sqrt{\langle \hat{P}^2 \rangle - \langle \hat{P} \rangle^2}$ are the position and momentum uncertainties, and $\beta$ is a small constant. This condition establishes the minimum possible value of the position uncertainty to be $\Delta X_{\text{min}} = \hbar \sqrt{\beta}$, which seems reasonable to define as the Planck length, so that $\beta \sim L_\text{P}^2/\hbar^2$. In the literature, the inequality of the above type is referred to as the generalized uncertainty principle (GUP). In order for the GUP to arise, the position and momentum operators must satisfy a modified commutation rule
\begin{equation}
	[\hat{X}, \hat{P}] = i \hbar \big( 1 + \beta \hat{P}^2 \big).
	\label{eq:GUP_commutator}
\end{equation}

The modification of the commutation relation has a profound impact on the formalism of quantum mechanics. In particular, the new theory encounters the absence of the familiar representation in the space of position wave functions. In fact, the eigenvectors of the position operator, which form the basis in the position representation, are found to lie outside the domain of physical states. This happens because these states, by definition, have zero position uncertainty, which is forbidden by the GUP. Conversely, the momentum operator eigenvectors are consistent with the GUP (more precisely, they do not actually belong to the Hilbert space, but we can approximate them arbitrarily well), so the momentum representation is still available for use.

It should be noted that \eqref{eq:GUP_commutator} is not the only deformation of the commutation rule that has been studied so far \cite{Nouicer2007, Nozari2012, Pedram2012}. Much attention has also been paid to the general form \cite{Detournay2002, Abdelkhalek2016}
\begin{equation}
	[\hat{X}, \hat{P}] = i \hbar f(\hat{P}), 
	\label{eq:f_commutator}
\end{equation}
where $f(p)$ is a smooth positive function that should incorporate the desired fundamental features into the theory. Specifically, $f(p)$ must grow at least linearly as $p \rightarrow \pm \infty$ to ensure a lower bound on position uncertainty. At the same time, it must depend on a parameter $\beta$ in such a way that $\lim_{\beta \rightarrow 0} f(p) = 1$ in order to be consistent with standard quantum mechanics.

As a further generalization of the above GUP, one can add the similar quadratic term for position uncertainty~\cite{Kempf1994}:
\begin{equation}
	\Delta X \Delta P \ge \frac{\hbar}{2} \big( 1 + \alpha \Delta X^2 + \beta \Delta P^2 \big).
	\label{eq:EGUP}
\end{equation}
This inequality, usually called the extended generalized uncertainty principle (EGUP), states that the uncertainties in both position and momentum can never be reduced to zero. To provide the EGUP, the commutation rule that underlie the theory should be postulated as
\begin{equation}
	[\hat{X}, \hat{P}] = i \hbar \big( 1 + \alpha \hat{X}^2 + \beta \hat{P}^2 \big).
	\label{eq:EGUP_commutator}
\end{equation}
An argument why the minimal momentum uncertainty should be introduced comes from considering the non-flat global structure of the universe. Indeed, the non-zero cosmological constant $\Lambda$ introduces a general curvature into space, making the plane waves with an infinitely narrow momentum distribution unrealizable. In this context, the coefficient $\alpha$, which determines the resolution in momentum, is taken to be $\alpha \sim \Lambda$ \cite{Mignemi2010}. In the result, we lose the momentum representation as well, so the only possible way to develop the formalism is to use the Fock representation. It is the theory that gave rise to $q$-deformed quantum mechanics for the case $q>1$ \cite{Hinrichsen1996}.

In the following, we start with the momentum-dependent commutation rule of the general form \eqref{eq:f_commutator}. Next, we will briefly cover the case \eqref{eq:EGUP_commutator} where the commutation rule depends on both position and momentum operators.

\section{Representation of the operators algebra}

\subsection{Deformation of operator algebra}

Our starting point is ordinary quantum mechanics, which is based on the Heisenberg commutation relation
\begin{equation}
	[\hat{x}, \hat{p}] = i \hbar \\
	\label{eq:HCR}
\end{equation}
between the operators $\hat{x}$ and $\hat{p}$ associated with the position and the momentum of a particle. The Hilbert space in which they act can be realized as the space of square integrable functions $\psi(p)$ with unit measure, $L^2(\mathbb{R}, dp)$, the inner product in which is
\begin{equation}
	\begin{aligned}
		& \bra{\varphi} \ket{ \psi} = \int \varphi^*(p) \psi(p) dp, \\
		& \varphi(p), \psi(p) \in L^2(\mathbb{R}, dp). \\
	\end{aligned}
	\label{eq:usual_inner_prod}
\end{equation}
In the momentum representation, $\hat{x}$ and $\hat{p}$ act on the appropriate domain of functions (wave functions) as
\begin{equation}
	\begin{aligned}
		& \hat{x} = i \hbar \partial_p, \\
		& \hat{p} = p, \\
	\end{aligned}
	\label{eq:canonical_XP}
\end{equation}
where $\partial_p$ denotes the derivative. This formalism is symmetric to transformations of operators and vectors of the form
\begin{equation}
	\begin{aligned}
		\hat{A} &\longrightarrow \hat{A}_U = \hat{U} A \hat{U}^{-1}, \\
		\ket{\varphi} &\longrightarrow \ket{\varphi_U} = \hat{U} \ket{\varphi}, \\
	\end{aligned}
	\label{eq:unitary_transformation}
\end{equation}
if the operator $\hat{U}$ is unitary ($\hat{U}^{-1} = \hat{U}^\dag$). Indeed, this transformation doesn't change the commutator \eqref{eq:HCR} and thus the operator algebra. At the same time, the expression for expectation values is invariant: $\bra{\varphi} \hat{A} \ket{\psi} = \bra{\varphi_U} \hat{A}_U \ket{\psi_U}$.

Let us trace the transition to the formalism, which involves the deformed commutation relation \eqref{eq:f_commutator}. The main constraint we impose is to preserve the unit measure of the Hilbert space. 

One approach involves deforming the momentum operator while leaving the position operator essentially unchanged \cite{Abdelkhalek2016}. For this, introduce the new position and momentum operators $\hat{X}$ and $\hat{P}$ by
\begin{equation}
	\begin{aligned}
		& \hat{x} = \hat{X}, \\
		& \hat{p} = k(\hat{P}),~~~ k(p) = \int\limits_{0}^{p} \frac{d \tilde{p}}{f(\tilde{p})}, \\
	\end{aligned}
	\label{eq:P_deform_approach}
\end{equation}
whose commutator is precisely \eqref{eq:GUP_commutator}. For example, in the specific case of $f(p) = 1 + \beta p^2$, the deformed momentum operator is $\hat{P} = \tan( \sqrt{\beta} \hat{p}) / \sqrt{\beta}$. In the momentum representation, the measure used in the inner product integration is unit as we were aiming for. However, the momentum set over which the integration is performed is bounded by the cutoff $|p| \le p_{max} = k(p \rightarrow +\infty)$ (which should be a finite value). This can be interpreted as restricting the set of physical states to those whose wave functions have support limited by the cutoff. The latter allows us to extend the inner product integration to the entire real line. Actually, we have arrived at the usual quantum formalism, which applies only to the states with truncated momentum. Such truncation is essentially a mechanism for excluding highly localized states that necessarily have too broad momentum distribution.

As an alternative, the position operator could be deformed instead. To do this in the most intuitive way, just formally multiply the commutator \eqref{eq:HCR} on the left side by an operator $f(\hat{p})$, where $f(p)$ is a positive function that satisfies the conditions previously discussed. This leads to the commutation relation of the form \eqref{eq:GUP_commutator} between the operators
\begin{equation}
	\begin{aligned}
		& \hat{X} = f(\hat{p}) \hat{x}, \\
		& \hat{P} = \hat{p}. \\
	\end{aligned}
	\label{eq:old_representation}
\end{equation}
However, the operator $\hat{X}$ is not self-adjoint or even symmetric within the original Hilbert space $L^2(\mathbb{R}, dp)$ because of $\hat{X}^\dag = \hat{x} f(\hat{p}) \ne \hat{X}$. As a consequence, the expectation values of $\hat{X}$ are complex, and therefore this operator cannot be considered as an observable. 
To ensure real expectation values, we can make $\hat{X}$ generally symmetric by introducing a new Hilbert space $L^2(\mathbb{R}, dp/f(p))$, which differs from the one used before by the measure $1/f(p)$ in the scalar product:
\begin{equation}
	\begin{aligned}
		& \bra{\Phi} \ket{\Psi} = \int \Phi^*(p) \Psi(p) \frac{dp}{f(p)},\\
		& \Phi(p), \Psi(p) \in L^2 \left( \mathbb{R}, \frac{dp}{f(p)} \right). \\
	\end{aligned}
\end{equation}
It is the pair of position and momentum operators \eqref{eq:old_representation} that is commonly used in the literature \cite{Kempf1995}. Note that the transition to a new variable $\rho$ given by $d\rho = dp/f(p)$ shows that the two paths above are essentially the same.

The natural question that now arises is whether the algebra of operators $\hat{X}$ and $\hat{P}$ is invariant under unitary transformations  $\hat{U} \thinspace \cdot \thinspace \hat{U}^\dag$, as it is in ordinary quantum mechanics, where $\hat{U}$ is a unitary operator. It is seen from \eqref{eq:f_commutator}, that such transformations generally change the right-hand side of the commutator. As a result, the minimum value of position uncertainty will vary, for example, for translated or rotated frames. To avoid this, it is natural to make the deformed commutator immune to unitary transformations by requiring $\hat{U} f(\hat{P}) \hat{U}^\dag = f(\hat{U} \hat{P} \hat{U}^\dag)$. The most practical way to ensure this is to take a polynomial function as $f(p)$.

\subsection{Non-unitary symmetry}

The above representation with position deformation can be transferred to the usual Hilbert space $L^2(\mathbb{R}, dp)$ using a non-unitary transformation. For this purpose, apply the transformation~\eqref{eq:unitary_transformation}, but using a non-unitary operator $\hat{V} \equiv V(\hat{P})$ that depends on only the momentum operator:
\begin{equation}
	\begin{aligned}
		& \hat{\mathcal{X}} = \hat{V} \hat{X} \hat{V}^{-1}, \\
		& \hat{\mathcal{P}} = \hat{V} \hat{P} \hat{V}^{-1} = \hat{P}, \\
	\end{aligned}
	\label{eq:xp_transformation}
\end{equation}
where $V(p)$ is a positive function. It is easy to see that the new position and momentum operators obey the deformed commutation relation \eqref{eq:f_commutator}. Therefore, the structure of the operator algebra has not been changed. However, the expression for expectation values in the transformed terms becomes
\begin{equation}
	\bra{\varphi} \hat{A} \ket{\psi} = \bra{\varphi_V} (\hat{V} \hat{V}^\dag)^{-1} \hat{A}_V \ket{\psi_V},
\end{equation}
despite the values themselves obviously remain constant. To bring this expression into an invariant form, one should compensate for the extra operator $(\hat{V} \hat{V}^\dag)^{-1}$ by introducing a suitable measure of integration into the inner product. A modification of the inner product, in turn, actually changes the Hilbert space. It is the freedom we use to further transform the representation of the deformed operators in order to incorporate them into the ordinary framework of quantum mechanics.

Let's apply the above transformation to the GUP-deformed momentum representation. By combining equations \eqref{eq:canonical_XP} and \eqref{eq:old_representation} and  substituting them into \eqref{eq:xp_transformation}, we deduce that in the momentum representation
\begin{equation}
	\begin{aligned}
		& \hat{\mathcal{X}} = i \hbar f(p) V \partial_p V^{-1} + i \hbar f(p) \partial_p, \\
		& \hat{\mathcal{P}} = p. \\
	\end{aligned}
\end{equation}
In this case, the measure of integration in the inner product should also be converted according to
\begin{equation}
	\frac{1}{f(p)} \longrightarrow \frac{1}{V^2 f(p)}. \\
	\label{eq:measure_transform}
\end{equation}
Indeed, only then will the position operator have the same expectation values (in particular, remain symmetric):
\begin{equation}
	\begin{aligned}
		& \bra{\psi} \Big( \hat{\mathcal{X}} \ket{\varphi} \Big) = \int \psi^*(p) \Big[ V  \hat{X} V^{-1} \varphi(p) \Big] \frac{dp}{V^2f(p)} = \\
		& i \hbar V^{-2} \psi^*(p) \varphi(p) \Big|_{-\infty}^{\infty} +  \int \Big[ V \hat{X} V^{-1} \psi(p) \Big]^* \varphi(p) \frac{dp}{V^2 f(p)} = \Big( \bra{\psi} \hat{\mathcal{X}}^\dag \Big) \ket{\varphi}. \\
	\end{aligned}
\end{equation}

It is clear from \eqref{eq:measure_transform} that the unit measure in the integration occurs when $V(p) = 1/\sqrt{f(p)}$ is chosen. The initial operators \eqref{eq:old_representation} are then transformed into
\begin{equation}
	\begin{aligned}
		& \hat{\mathcal{X}} = \sqrt{f(\hat{p})} \hat{x} \sqrt{f(\hat{p})}, \\
		& \hat{\mathcal{P}} = \hat{p}, \\
	\end{aligned}
	\label{eq:gen_XP_representation}
\end{equation}
which now act wave functions from the ordinary Hilbert space $L^2(\mathbb{R}, dp)$ as
\begin{equation}
	\begin{aligned}
		& \hat{\mathcal{X}} = i \hbar f(p) \partial_p + \frac{i \hbar}{2} f'(p),  \\
		& \hat{\mathcal{P}} = p. \\
	\end{aligned}
	\label{eq:XP_representation}
\end{equation}
The latter form clearly reveals the relationship between the new and original operators, which is $\hat{\mathcal{X}} = \hat{X} + i \hbar f'(\hat{P})/2$. We immediately notice that the operators $\hat{\mathcal{X}}$ and $\hat{\mathcal{P}}$ are symmetric from the point of view of the canonical formalism of quantum mechanics. Indeed, the straightforward verification shows that using the usual scalar product with unit measure \eqref{eq:usual_inner_prod} yields
\begin{equation}
	\begin{aligned}
		& \bra{\psi} \Big( \hat{\mathcal{X}} \ket{\varphi} \Big) = \int \psi^*(p) \Big[ i \hbar f(p) \partial_p + \frac{i \hbar}{2} f'(p) \Big] \varphi(p) dp = \\
		& = i \hbar \int \psi^*(p) f(p) \varphi(p) \Big|_{-\infty}^{\infty} + \int \left[ \left( i \hbar f(p) \partial_p + \frac{i \hbar}{2} f'(p) \right) \psi(p) \right]^* \varphi(p) dp = \Big( \bra{\psi} \hat{\mathcal{X}}^\dag \Big) \ket{\varphi}. \\
	\end{aligned}
\end{equation}
Importantly, the resulting operators \eqref{eq:XP_representation} do obey the generalized commutator relation \eqref{eq:f_commutator}.

It should be noticed that the deformed position operator $\hat{\mathcal{X}}$ admits another form
\begin{equation}
	\hat{\mathcal{X}} = \hat{x} + \sqrt{g(\hat{p})} \hat{x} \sqrt{g(\hat{p})}, \\
	\label{eq:X_through_g}
\end{equation}
where $f(p) = 1 + g(p)$. Indeed, the straightforward calculation reveals that both \eqref{eq:gen_XP_representation} and \eqref{eq:X_through_g} have exactly the same representation \eqref{eq:XP_representation} in the space of momentum wave functions. However, this form of the position operator is more convenient in certain cases, since it explicitly demonstrates the deviation of the deformed operator from the canonical one. As an example, using \eqref{eq:X_through_g} makes it more transparent to determine appropriate states for which the uncertainty relation is satisfied. Specifically, the GUP holds for all states from the domain
\begin{equation}
	D(\hat{x}\hat{p}) \cap D(\hat{p}\hat{x}) \cap D( \sqrt{\hat{g}} \hat{x} \sqrt{\hat{g}} \hat{p}) \cap D(\hat{p} \sqrt{\hat{g}} \hat{x} \sqrt{\hat{g}}) \cap D(\hat{g}) \\
	\label{eq:GUP_domain}
\end{equation}
where we denoted $g(\hat{p})$ as $\hat{g}$. It is explicitly seen that this domain is a subdomain of
\begin{equation}
	D(\hat{x}\hat{p}) \cap D(\hat{p}\hat{x}), \\
\end{equation}
for states from which the Heisenberg uncertainty principle holds \cite{Hall2013}. However, these peculiarities can be addressed by manually defining the domain of physical states as the Schwartz space $\mathcal{S}$, that is, the space of all functions that, together with all their derivatives, decrease with $p$ faster than any power of $p$. In doing so, we actually use the common domain of physical states for both the canonical and deformed formalisms.


\subsection{The position operator}

The deformation of the position operator introduces crucial properties that are significant for further analysis. It has been shown that in the self-consistent formalism with the Hilbert space $L^2(\mathbb{R}, dp/f(p))$, the operator $\hat{X}$ (see \eqref{eq:old_representation}) is essentially self-adjoint and has a family of self-adjoint extensions \cite{Kempf1995}. However, its eigenvectors are unphysical because they neither obey the GUP themselves, nor can they be approximated by vectors obeying the GUP. By the same arguments, the behavior of $\hat{\mathcal{X}}$ acting in $L^2(\mathbb{R}, dp)$ is similar. The eigenvector problem
\begin{equation}
	\hat{\mathcal{X}} \varphi_x(p) = x \varphi_x(p) \\
\end{equation}
results in 
\begin{equation}
	\varphi_x(p) = \left( \int\limits_{-\infty}^{+\infty} \frac{d \tilde{p}}{f(\tilde{p})} \right)^{-\frac{1}{2}} \frac{1}{\sqrt{f(p)}} \exp\left\{ -\frac{i}{\hbar} x \int\limits_0^p \frac{d \tilde{p}}{f(\tilde{p})} \right\}. \\
	\label{eq:defX_eigenstate}
\end{equation}
These vectors inherently possess zero uncertainty and thus cannot be physical.
Among the physical states, the most localized is the one that exhibits the least possible uncertainty in position. It can be derived by setting the conditions for the specific mean value for position
\begin{equation}
	\bra{\varphi^{ML}_\xi} \hat{\mathcal{X}} \ket{\varphi^{ML}_\xi} = \xi \\
\end{equation}
and for uncertainty $\Delta \mathcal{X} \rightarrow \min$. It is the state that is referred to as the minimum-length state. Following~\cite{Kempf1995}, its wave function can be found from the equality
\begin{equation}
	\Big( \hat{\mathcal{X}} - \xi + \frac{\langle [\hat{\mathcal{X}}, \hat{\mathcal{P}}] \rangle}{2 (\Delta \mathcal{P})^2} (\hat{\mathcal{P}} - \eta)\Big) \varphi^{ML}_\xi(p) = 0, \\
	\label{eq:gen_min_uncert_state_problem}
\end{equation}
where $\eta$ is the average value of the momentum. In the general case, however, this problem is challenging because the mean value of the commutator here depends on the state itself and can only be solved for a few specific functions $f(p)$ \cite{Abdelkhalek2016}. In the most popular case of $f(p) = 1 + \beta p^2$, the problem becomes simpler and yields
\begin{equation}
	\varphi^{ML}_\xi(p) = \sqrt{\frac{2\sqrt{\beta}}{\pi}} \frac{1}{1 + \beta p^2} \exp \left\{ - i\frac{\xi}{\hbar \sqrt{\beta}} \atan \sqrt{\beta} p \right\}. \\
	\label{eq:ML_wave_function}
\end{equation}
We see, that both wave functions \eqref{eq:defX_eigenstate} and \eqref{eq:ML_wave_function} differ from those derived in \cite{Kempf1995} for the Hilbert space $L^2(\mathbb{R}, dp/f(p))$ by a factor $1/\sqrt{1 + \beta p^2}$, which arises due to the change of measure in the scalar product.

It is also interesting to note that from \eqref{eq:gen_min_uncert_state_problem} it directly follows that the eigenvectors of an operator $\hat{\xi}$, defined as
\begin{equation}
	\hat{\xi} \varphi^{ML}_\xi(p) = \left( \hat{\mathcal{X}} + i  \hbar \beta \hat{\mathcal{P}} \right) \varphi^{ML}_\xi(p) = \xi \varphi^{ML}_\xi(p), \\
\end{equation}
are the very minimum-length states \eqref{eq:ML_wave_function}. 

Since here we consider the deformed position operator $\hat{\mathcal{X}}$ as an operator from the ordinary framework of quantum mechanics with the Hilbert space $L^2(\mathbb{R}, dp)$, it is instructive to examine the canonical position basis from the point of view of the GUP. For this, consider a particular state with the Gaussian form
\begin{equation}
	\psi_s(p) = \left( \frac{a}{\pi} \right)^{\frac{1}{4}} e^{-\frac{a}{2} (p - p_0)^2 - \frac{i}{\hbar} p x_0}, \\
	\label{eq:squeezed_state}
\end{equation}
where $a > 0$ is the squeezing parameter. The expectation values of the canonical position and momentum in this state  are
\begin{equation}
	\langle \hat{x} \rangle = x_0, ~~~\langle \hat{p} \rangle = p_0,
\end{equation}
and their uncertainties are
\begin{equation}
	\Delta x = \sqrt{\frac{\hbar^2 a}{2}}, ~~~\Delta p = \sqrt{\frac{1}{2a}}. \\
	\label{eq:canonical_uncertainties}
\end{equation}
Both uncertainties vary oppositely with $a$ in full accordance with the Heisenberg uncertainty principle and saturate it. The important point for us here is that the state $\ket{\psi_s}$ tends to $\ket{x_0}$ in the limit $a \rightarrow 0$ and to $\ket{p_0}$ in the limit $a \rightarrow +\infty$.

To understand the characteristics of the state $\ket{\psi_s}$ in the deformed case, the first thing to note is that its wave function belongs to the Schwartz space $\mathcal{S}$. This implies that $\psi_s(p)$ is within the joint domain of all operators appearing in the deformed commutation relation, so the GUP definitely holds for any $a$. Consequently, the canonical position eigenstates $\ket{x}$ are suitable for the GUP theory in their limit sense. While the momentum uncertainty $\Delta \mathcal{P}$ for the state $\ket{\psi_s}$ is obviously the same as in \eqref{eq:canonical_uncertainties}, the deformed position uncertainty is
\begin{equation}
	\begin{aligned}
		\Delta \mathcal{X} = \Bigg[ \frac{\hbar^2 a}{2} \left( 1 + \frac{\beta}{a} + \frac{7}{4} \frac{\beta^2}{a^2} \right) +\hbar^2 a \beta p_0^2 \left( 1 + \frac{1}{2} \beta p_0^2 \right) + \frac{5}{2} \hbar^2 \beta^2 p_0^2 + \frac{2\beta^2}{a} x_0^2 p_0^2 + \frac{\beta^2}{2a^2} x_0^2 \Bigg]^{\frac{1}{2}}. \\
	\end{aligned}
\end{equation}
From this relation, we see that uncertainty in the deformed position increases indefinitely in the limit $a \rightarrow\infty$, similar to its canonical counterpart. But unlike the latter, $\Delta \mathcal{X}$ also diverges in the limit $a \rightarrow 0$. This happens due to the additional terms that are inversely dependent on~$a$. 

In result, both the momentum and position uncertainties of the state $\ket{\psi_s}$ diverge when $a \rightarrow 0$. In other words, we know totally nothing about neither the particle's position, nor its momentum. Since $\ket{\psi_s}$ approaches $\ket{x_0}$ in this limit, we conclude that the canonical position eigenstates lose any physical sense within the GUP formalism, although they can still be used for calculations. But in order to have a physical meaning, the results of calculations should be expressed in terms of the momentum basis only.

Between the above limits, there exists a positive value of $a$ at which uncertainty in the deformed position of the state $\ket{\psi_s}$ reaches a non-zero minimum value. This is nothing but a manifestation of the "minimum-length" property of the theory. To demonstrate this, let's vary $a$ from infinity to zero. In the canonical formalism, we start from the state with completely unknown localization and proceed sequentially through states that are more and more localized until we reach the precisely localized state. In the deformed case, if we go through the same set of the states $\ket{\psi_s}$ by decreasing $a$, $\Delta \mathcal{X}$ also initially starts from infinity. But there is a threshold value of $a$, beyond which the uncertainty turns and goes to infinity again. The states considered are not those that saturate the GUP, as they really do for the Heisenberg uncertainty principle. Nevertheless, this example provides a qualitative demonstration of the underlying idea of the deformed formalism.


\subsection{Three-dimensional case}

The extension to the three-dimensional case where uncertainty in each deformed position has a non-zero minimum is straightforward:
\begin{equation}
	[\hat{\mathcal{X}}_i, \hat{\mathcal{P}}_j] = i \hbar \delta_{ij} f(|\hat{\bm{\mathcal{P}}}|).
\end{equation}
Here, the right-hand side depends on the momentum modulus $|\hat{\bm{\mathcal{P}}}| = \sqrt{\sum_i \hat{\mathcal{P}}_i^2}$ in order to preserve rotational symmetry. The position and momentum operators obeying these commutation relations now become
\begin{equation}
	\begin{aligned}
		& \hat{\mathcal{X}}_i = \sqrt{f(|\hat{\bold{p}}|)} \hat{x}_i \sqrt{f(|\hat{\bold{p}}|)} \\
		& \hat{\mathcal{P}}_i = \hat{p}_i \\
	\end{aligned}
\end{equation}
Since the deformed position operators mix the canonical position and momentum operators, they are no longer commute \cite{Kempf1995}:
\begin{equation}
	[\hat{\mathcal{X}}_i, \hat{\mathcal{X}}_j] = i \hbar \hat{f}^{\frac{1}{2}} \left( \partial_{p_i} \hat{f} \hat{\mathcal{X}}_j - \partial_{p_j} \hat{f} \hat{\mathcal{X}}_i \right) \hat{f}^{-\frac{1}{2}}, \\
\end{equation}
where $\hat{f} \equiv f(|\hat{\bm{\mathcal{P}}}|)$. This means that no pair of the deformed coordinates can be precisely measured.

It is important to note the conceptual difference between the self-consistent framework of the GUP-generating operator algebra \cite{Kempf1994} and that which is inseparable from the canonical quantum framework, as considered here. In the present interpretation, the fundamental properties of the ordinary operator algebra still hold; we just regard the deformed position operator as an operator embedded within the canonical algebra. Consequently, all features of the deformed algebra are determined by the canonical framework; there is no necessity for additional assumptions about the form of the deformed operator algebra. For example, the commutativity of the deformed momentum operators $[\hat{\mathcal{P}}_i, \hat{\mathcal{P}}_j] = 0$ directly stems from the canonical algebra. Moreover, no specific requirements for the mutual commutators of the coordinates are necessary as well \cite{Fadel2022}.


\section{The deformed oscillator}

In this section, consider the energy shift that occurs in a deformed quantum oscillator. Let the Hamiltonian of the deformed system be
\begin{equation}
	\hat{H}_{GUP} = \hbar \omega \left( \frac{\hat{\mathcal{P}}^2}{4 K_0^2} + \frac{\hat{\mathcal{X}}^2}{4 L_0^2} \right), \\
	\label{eq:deformed_osc_H}
\end{equation}
where
\begin{equation}
	L_0 = \sqrt{\frac{\hbar}{2 m \omega}},~~ K_0 = \sqrt{\frac{\hbar m \omega}{2}}. \\
	\label{eq:L0K0}
\end{equation}
Substituting the deformed position operator \eqref{eq:X_through_g} here, the problem is reduced to the ordinary quantum oscillator with the usual Hamiltonian
\begin{equation}
	\hat{H}_0 = \hbar \omega \left( \frac{\hat{p}^2}{4 K_0^2} + \frac{\hat{x}^2}{4 L_0^2} \right), \\
	\label{eq:harmonic_osc_H}
\end{equation}
which is placed in the special potential $\hat{V}$:
\begin{equation}
	\hat{V} = \frac{\hbar \omega}{4 L_0^2} \left( \hat{x} \sqrt{\hat{g}} \hat{x} \sqrt{\hat{g}} + \sqrt{\hat{g}} \hat{x} \sqrt{\hat{g}} \hat{x} + \sqrt{\hat{g}} \hat{x} \hat{g} \hat{x} \sqrt{\hat{g}} \right), \\
\end{equation}
where we have denoted $\hat{g} \equiv g(\hat{p})$ for brevity. In the momentum representation, $\hat{V}$ acts wave functions $\psi(p)$ as
\begin{equation}
	\begin{aligned}
		\hat{V} \psi =  - \frac{\hbar^3 \omega}{4 L_0^2} &\left[ g(2 + g) \psi'' + 2(1 + g) g' \psi' + \frac{1}{2} \left(g g'' + \frac{g'^2}{2} + g'' \right) \psi \right]. \\
	\end{aligned}
\end{equation}

Given that the function $g(p)$ depends on the small parameter $\beta$, we can further apply the perturbation approach. 
Let the wave function and the energy associated with it be $\psi_n = \psi_n^0 + \delta \psi_n$ and $E_n = E_n^0 + \delta E_n$, where
\begin{equation}
	\begin{aligned}
		& \psi_n^0(p) = \frac{1}{\sqrt{2^n n! \sqrt{\pi}}} \left( \frac{1}{2K_0^2} \right)^{1/4} H_n \left(  \frac{p}{\sqrt{2} K_0} \right) e^{-\frac{p^2}{4 K_0^2} }, \\
		& E_n^0 = \hbar \omega \left(n + \frac{1}{2} \right)
	\end{aligned}
\end{equation}
are the solutions and the corresponding energies for an unperturbed harmonic oscillator. Then, the first-order correction $\delta \psi$ can be found by solving the differential equation
\begin{equation}
	(\hat{H}^0 - E_n^0) \delta \psi_n = - (\hat{V} - \delta E_n) \psi_n^0  \\
\end{equation}
where
\begin{equation}
	\delta E_n = \int {\psi_n^0}^* \hat{V} \psi_n^0 dp \\
\end{equation}
is the correction to the energy of the harmonics oscillator. For the case $g = \beta p^2$, the energy corrections are
\begin{equation}
	\delta E_n = \frac{\beta m \hbar^2 \omega^2}{2} \left(n^2 + n + \frac{1}{2} \right), \\
\end{equation}
which match with presented in \cite{Kempf1995}.

\section{Wigner function within deformed formalism}

The study of the Wigner representation in the context of deformed quantum mechanics also attracts considerable interest. The advantage of this representation is that it closely resembles the formalism of statistical mechanics in phase space. The background concept here is the Weyl transformation, which maps an operator $\hat{A}$ acting in the Hilbert space into a $c$-function $\widetilde{A}(x,p)$ defined in the phase space \cite{Case2008}
\begin{equation}
	\widetilde{A}(x,p) = \int e^{\frac{i}{\hbar} x u} \bra{p + \frac{u}{2}} \hat{A} \ket{p - \frac{u}{2}} du.
	\label{eq:Weyl_transformation}
\end{equation}
When the transformation is applied to the density matrix representing a quantum state, the Wigner function is obtained:
\begin{equation}
	W(x,p) = \frac{2\pi}{\hbar} \widetilde{\rho}(x,p).
\end{equation}
It is well known, that the Wigner function is not a true probability distribution because it can take negative values. However, integrating the Wigner function over $x$ or $p$ yields the probability distribution of momentum or position, respectively:
\begin{equation}
	\begin{aligned}
		& \int W(x,p) dx = \bra{p} \hat{\rho} \ket{p}, \\
		& \int W(x,p) dp = \bra{x} \hat{\rho} \ket{x}. \\
	\end{aligned}
	\label{eq:marginals}
\end{equation}

The crucial point is that the Weyl transformation is isomorphic. This stems from the orthogonality and completeness of the displacement operators $\hat{D}(u,v) = e^{-\frac{i}{\hbar} (u \hat{x} + v \hat{p}) }$, which are used to construct the transformation \cite{Mukunda2005}:
\begin{equation}
	\Tr\{ \hat{D}(u_1,v_1) \hat{D}^\dag(u_2,v_2) \} = 2\pi \hbar \delta(u_2 - u_1) \delta(v_2-v_1). \\
\end{equation}
It is the property that allows the Weyl transform to be used for calculating the expectation values of quantum observables in a manner similar to classical statistical mechanics:
\begin{equation}
	\langle A \rangle = \Tr \{\hat{A} \hat{\rho} \} = \int \widetilde{A}(x,p) W(x,p) dx dp. \\
	\label{eq:average_using_WF}
\end{equation}

The Weyl transform within the self-consistent theory with the GUP is hardly available. Indeed, for property~\eqref{eq:average_using_WF} to be applied in the deformed theory, the existence of a set of orthogonal operators, analogous to the displacement operators in the canonical formalism, is required. The main problem here is the lack of such a set. The naive replacement of the position and momentum operators $\hat{x}$ and $\hat{p}$ with their GUP-deformed counterparts $\hat{\mathcal{X}}$ and $\hat{\mathcal{P}}$ does not yield the desired result: we cannot even detach them in the exponential due to their non-trivial commutator. 

However, the present approach to regard the GUP while staying within the Hilbert space of ordinary quantum mechanics $L^2(\mathbb{R}, dp)$ might be a way around this problem. Specifically, one can continue using the displacement operator $\hat{D}$ in its canonical form. 
This means that the Wigner function is calculated using the usual Weyl transformation \eqref{eq:Weyl_transformation}. Then we preserve properties like~\eqref{eq:average_using_WF} and marginal distributions \eqref{eq:marginals}. The only subtlety here comes from the fact that the position eigenvectors, and thus the position distribution in \eqref{eq:marginals}, lose their physical sense. Instead, this relationship should be rewritten in terms of the momentum basis:
\begin{equation}
	\int W(x,p) dp = \frac{1}{2\pi} \int d\eta d\eta' e^{i (\eta - \eta') x} \bra{\eta} \hat{\rho} \ket{\eta'}. \\
\end{equation}

Now consider the Weyl transform of the operators corresponding to the basic observables. The transform of the unit and the momentum operator, obviously, remains the same: $\widetilde{1}(x, p) = 1$ and $\widetilde{\mathcal{P}}(x, p) = p$. As for the deformed position operator, its transform is more complicated:
\begin{equation}
	\widetilde{\mathcal{X}}(x,p) = x f(p), \\
\end{equation}
which corresponds to \eqref{eq:gen_XP_representation} when the latter is considered as a simple variable. However, the transformation of the squared position operator 
\begin{equation}
	\widetilde{\mathcal{X}^2}(x,p) = x^2 f^2(p) + \hbar^2 \frac{f'^2(p)}{4} \\
\end{equation}
already contains an additional term occurring due to non-trivial structure of the operator $\hat{\mathcal{X}}$.

To conclude this section, it is worth noting that the GUP-deformed Weyl transformation proposed in \cite{Yeole2021} is actually the canonical one \eqref{eq:Weyl_transformation}. Indeed, both do coincide if the wave functions $\Psi$ used in \cite{Yeole2021}, which actually live in the Hilbert space $L^2(\mathbb{R}, dp/f(p))$, are transferred to the ordinary Hilbert space by the non-unitary transformation \eqref{eq:xp_transformation}: $\psi = \Psi/\sqrt{f(p)}$.

\section{The extended generalized uncertainty principle}

In this section, we address the more general case in which both position and momentum exhibit non-zero minimum uncertainty. Formally, this is described by the EGUP \eqref{eq:EGUP}, which follows from the commutation relation~\eqref{eq:EGUP_commutator} involving both momentum and position squared terms. The most significant aspect that arises here is that although the eigenvectors of the position and momentum operators do occur, they violate the EGUP due to their inherent zero uncertainty. This implies that these vectors do not correspond to any physical state, and therefore neither the position nor the momentum representations are available. For this reason, it seems natural to use the Fock representation instead. It turns out, however, that such a symmetry between the operators makes it somewhat easier to treat the deformed formalism as incorporated into the canonical one.

In the Fock representation, the canonical position and momentum operators are expressed as
\begin{equation}
	\begin{aligned}
		& \hat{x} = L_0 (\hat{b}^\dag + \hat{b}), \\
		& \hat{p} = i K_0 (\hat{b}^\dag - \hat{b}), \\
	\end{aligned}
	\label{eq:xp_through_ladders}
\end{equation}
where $L_0$ and $K_0$ are constants defined in \eqref{eq:L0K0}, and $\hat{b}^\dag$ and $\hat{b}$ are the bosonic creation and annihilation (ladder) operators which increase and decrease the number of particles in number states:
\begin{equation}
	\begin{aligned}
		& \hat{b}^\dag \ket{n}_b = \sqrt{n + 1} \ket{n + 1}_b, \\
		& \hat{b} \ket{n}_b = \sqrt{n} \ket{n - 1}_b.
	\end{aligned}
	\label{eq:b}
\end{equation}
In these terms, the canonical commutator relation \eqref{eq:HCR} takes the well-known form $[\hat{b}, \hat{b}^\dag] = 1$.

Now define $q$-deformed ladder operators as \cite{Dey2015}
\begin{equation}
	\begin{aligned}
		& \hat{a} = F(\hat{N}_b) \hat{b}, \\
		& \hat{a}^\dag = \hat{b}^\dag F(\hat{N}_b), \\
	\end{aligned}
\end{equation}
where $\hat{N}_b = \hat{b}^\dag \hat{b}$ is the canonical number operator, the function $F(n)$ is defined as
\begin{equation}
	F(n) = \sqrt{\frac{[n+1]}{n+1}}, \\
\end{equation}
and $[n]$ is the $q$-number of $n$:
\begin{equation}
	[n] = \frac{q^n - 1}{q - 1}. \\
\end{equation}
Using \eqref{eq:b} we see, that the operators $\hat{a}$ and $\hat{a}^\dag$ act the number states as
\begin{equation}
	\begin{aligned}
		& \hat{a}^\dag \ket{n} = \sqrt{[n + 1]} \ket{n + 1},  \\
		& \hat{a} \ket{n} = \sqrt{[n]} \ket{n - 1}, 
	\end{aligned}
	\label{eq:a}
\end{equation}
so they indeed generate the ladder structure in the Fock space. These operators, however, form the $q$-deformed commutation relation \cite{Kempf1994}
\begin{equation}
	\hat{a} \hat{a}^\dag - q \hat{a}^\dag \hat{a} = 1. \\
	\label{eq:a_commutator}
\end{equation}
It is the relation that gives non-zero minimal uncertainties in both position and momentum. Indeed, define $q$-deformed position and momentum in the usual way, but using generalized ladder operators, as
\begin{equation}
	\begin{aligned}
		& \hat{\mathcal{X}} \equiv L(\hat{a}^\dag + \hat{a}) = L \left[ \hat{b}^\dag F(\hat{N}_b) + F(\hat{N}_b) \hat{b} \right], \\
		& \hat{\mathcal{P}} \equiv iK(\hat{a}^\dag - \hat{a}) = iK \left[ \hat{b}^\dag F(\hat{N}_b) - F(\hat{N}_b) \hat{b} \right]. \\
	\end{aligned}
	\label{eq:XP_EGUP_general}
\end{equation}
Then, relation \eqref{eq:a_commutator}, expressed in these terms, becomes \eqref{eq:EGUP_commutator} for $q>1$. Here, $q$ is defined from the equality $4KL=\hbar(1+q)$, and the parameters $L$ and $K$ establish minimum values for the position and momentum uncertainties through $\Delta \mathcal{X}_{\text{min}} = L \sqrt{(q-1)/q}$ and $\Delta \mathcal{P}_{\text{min}} = K \sqrt{(q-1)/q}$ \cite{Kempf1994}. Expressing the new position and momentum operators through the canonical ones, one gets
\begin{equation}
	\begin{aligned}
		& \hat{\mathcal{X}} = \frac{L}{2} \Big( \{f(\hat{N}_b), \hat{x}\} + i [f(\hat{N}_b), \hat{p}] \Big), \\
		& \hat{\mathcal{P}} = \frac{K}{2i} \Big( [f(\hat{N}_b), \hat{x}] + i \{f(\hat{N}_b), \hat{p}\} \Big), \\
	\end{aligned}
\end{equation}
where $\{ ~ , ~ \}$ is the anticommutator. In these expressions, the number operator should be expressed in terms of position and momentum as well:
\begin{equation}
	\hat{N}_b = \frac{\hat{x}^2}{4L_0^2} + \frac{\hat{p}^2}{4 K_0^2} - \frac{1}{2}. \\
\end{equation}

We see that both the canonical and the deformed ladder operator algebras share essentially the same Hilbert space. Therefore, the deformed quantum mechanics with non-zero limits of the position and momentum uncertainty emerges when the operators $\hat{\mathcal{X}}$ and $\hat{\mathcal{P}}$ are assigned to these observables instead of the canonical $\hat{x}$ and $\hat{p}$. 

The $q$-deformed quantum theory has been widely discussed in the literature and is formulated in a self-consistent on the basis of $q^{-1}$-Hermite polynomials \cite{Hinrichsen1996}. Here we will limit ourselves to the first-order approximation in the background of the canonical formalism. Specifically, represent $q = 1 + \epsilon$, where $0 < \epsilon \ll 1$ is a small parameter. The linear approximation of $\hat{a}$ and $\hat{a}^\dag$ in $\epsilon$ leads to non-linear deviations from the ordinary ladder operators
\begin{equation}
	\begin{aligned}
		& \hat{a} = \hat{b} + \frac{\epsilon}{4} \hat{b}^\dag \hat{b} \hat{b},  \\
		& \hat{a}^\dag = \hat{b}^\dag + \frac{\epsilon}{4} \hat{b}^\dag \hat{b}^\dag \hat{b}. \\
	\end{aligned}
\end{equation}
Substituting these into \eqref{eq:XP_EGUP_general}, one gets the deformed position and momentum operators
\begin{equation}
	\begin{aligned}
		& \hat{\mathcal{X}} = L \left[ \left(1 - \frac{\epsilon}{4} \right) \frac{\hat{x}}{L_0} + \frac{\epsilon}{16} \left( \frac{\hat{x}^3}{L_0^3} + \frac{\hat{p} \hat{x} \hat{p}}{K_0^2 L_0} \right) \right], \\
		& \hat{\mathcal{P}} = K \left[ \left(1 - \frac{\epsilon}{4} \right) \frac{\hat{p}}{K_0} + \frac{\epsilon}{16} \left( \frac{\hat{x} \hat{p} \hat{x}}{K_0 L_0^2} + \frac{\hat{p}^3}{K_0^3} \right) \right]. \\
	\end{aligned}
	\label{eq:XP_EGUP_approx}
\end{equation}
Using these definitions, it is instructive to examine their uncertainties in the squeezed state \eqref{eq:squeezed_state}, as previously discussed for the GUP. A direct derivation yields expressions of the form
\begin{equation}
	\begin{aligned}
		& \Delta \mathcal{X} = \left( \frac{\hbar^2 a}{2}  \sum_{n=-2}^2 B_n a^n \right)^{\frac{1}{2}}, \\
		& \Delta \mathcal{P} = \left( \frac{1}{2a} \sum_{n=-2}^2 A_n a^n \right)^{\frac{1}{2}}, \\
	\end{aligned}
\end{equation}
where $A_i$ and $B_i$ are constants. It can be observed that $\Delta \mathcal{X}$ and $\Delta \mathcal{P}$ go to infinity in both limits $a \rightarrow 0$ and $a \rightarrow +\infty$, in which the state $\ket{\psi_s}$ tends to the eigenvectors of the canonical position and momentum, $\ket{x}$ and $\ket{p}$. This implies that we lack any information about the deformed position and momentum in these states, which makes it impossible to assign a physical interpretation to them. Therefore, although the states remain in the physical domain (in the limit sense) and can still be used for calculations, the result should be expressed using the Fock basis in order to make it meaningful.

Let us also analyze the energy levels deviation of the deformed oscillator in the case of the EGUP. Inserting \eqref{eq:XP_EGUP_approx} into the Hamiltonian \eqref{eq:deformed_osc_H} gives
\begin{equation}
	\hat{H}_{GUP} = \hat{H}_0 + \frac{\epsilon}{2} \left( \frac{\hbar \omega}{4} - \hat{H}_0  + \frac{1}{\hbar \omega} \hat{H}_0^2 \right), \\
\end{equation}
where $\hat{H}_0$ is the Hamiltonian of a quantum harmonic oscillator \eqref{eq:harmonic_osc_H}. Using the perturbative approach, it immediately follows that the corrections to the energy levels will be
\begin{equation}
	\delta E_n = \frac{\epsilon}{2} \hbar \omega n^2.
\end{equation}

\section{Conclusion}

It is natural to expect that, in general, the particle's position will be affected by the gravitational influence of the measuring probe; the phenomenon being hardly noticeable before the Planck scale, but increases significantly beyond it. The generalization of the commutation rules in quantum mechanics provides a phenomenological mechanism for implementing such influence into the theory. Formally, it is realized by inserting momentum-dependent terms into the position-momentum commutation relations. The generalized uncertainty principle they induce leads to the absolute non-zero minimum in position uncertainty. 

The theory arising from the modified operator algebra turned out to be separate from ordinary quantum mechanics. In particular, one is forced to equip the Hilbert space with the scalar product of non-unitary measure in order to keep the position operator symmetric. This complicates the comparison of these two theories. In the present work, we have shown that the representation of the deformed operator algebra can be reformulated to stay within the canonical Hilbert space. This implies that it is sufficient just to introduce a specific operator responsible for the deformed position within the scope of the canonical quantum mechanics. Formulated in this way, quantum theory incorporating the generalized uncertainty principle operates with the same wave functions in the momentum representation to represent quantum states as the canonical formalism. The developed approach offers a novel perspective on the deformed quantum mechanics with non-zero minimal position uncertainty.

\bibliographystyle{ieeetr}
\bibliography{refs}

\end{document}